\documentclass[prl,onecolumn,noshowpacs,nobalancelastpage,floatfix]{revtex4}

\pdfoutput=1

\usepackage{amssymb}
\usepackage{graphicx}
\usepackage[dvips]{epsfig}
\usepackage{bm}
\usepackage{amsmath}
\usepackage{wrapfig}

\begin{document}

\title{The Quantum Internet}
\author{H. J. Kimble}
\affiliation{Norman Bridge Laboratory of Physics 12-33\\
California Institute of Technology\\
Pasadena, California 91125, USA}
\date{\today }

\begin{abstract}
Quantum networks offer a unifying set of opportunities and challenges across
exciting intellectual and technical frontiers, including for quantum
computation, communication, and metrology. The realization of quantum
networks composed of many nodes and channels requires new scientific
capabilities for the generation and characterization of quantum coherence
and entanglement. Fundamental to this endeavor are quantum interconnects
that convert quantum states from one physical system to those of another in
a reversible fashion. Such quantum connectivity for networks can be achieved
by optical interactions of single photons and atoms, thereby enabling
entanglement distribution and quantum teleportation between nodes.

\end{abstract}

\maketitle

\section{Introduction}

The past two decades have witnessed a wide range of fundamental discoveries
in quantum information science (QIS), ranging from a quantum algorithm that
places public-key cryptography at risk to a protocol for the teleportation
of quantum states \cite{nielsen-chuang}. This union of quantum mechanics and
information science has fostered great advances in our understanding of the
quantum world and in our ability to control coherently individual quantum
systems \cite{zoller05}. Unique capabilities of quantum systems to process
and distribute information have been identified, and powerful new
perspectives for understanding the complexity and subtleties of quantum
dynamical phenomena have emerged.

\begin{figure}[h]
\begin{center}
\includegraphics[width=13cm]{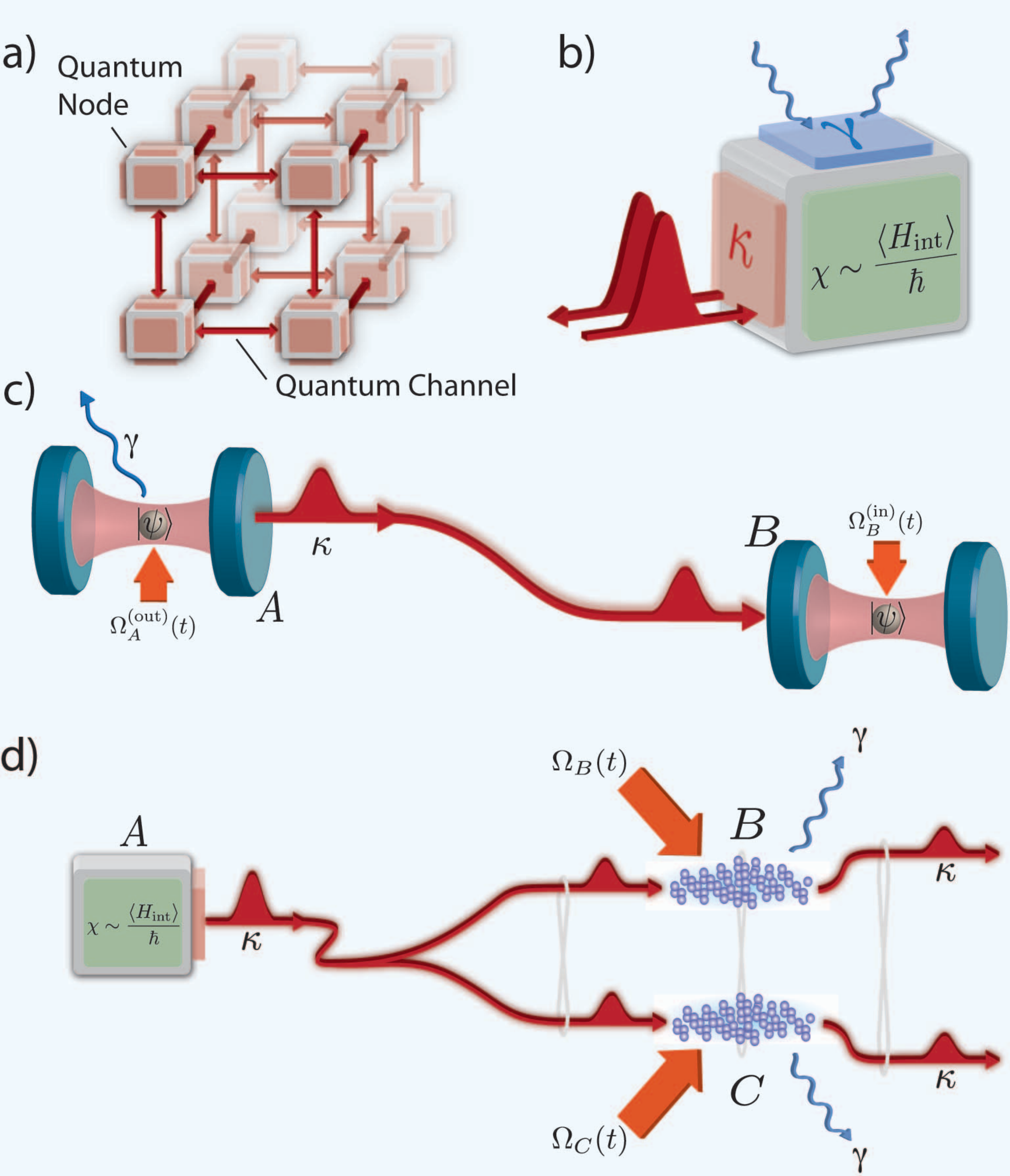}
\end{center}
\caption{Quantum networks from abstract to physical. (a) A quantum network
composed of \textit{quantum nodes} for processing and storing quantum states
and \textit{quantum channels} for the distribution of quantum information.
Such a network can alternatively be viewed as a strongly correlated
many-particle system. (b) Quantum interface between matter and light.
Coherent interactions within the node are characterized by the rate $\protect%
\chi $, while $\protect\kappa $ specifies the rate for coupling between the
node and photons in the external channel. Parasitic losses occur at rate $%
\protect\gamma $. (c) Quantum state transfer and entanglement distribution
from node \textit{A} to \textit{B} within the setting of cavity QED
\protect\cite{cirac97}. At node $A$ the control pulse $\Omega _{A}^{out}(t)$
affects the transformation of atomic state $\left\vert \protect\psi %
\right\rangle $ to the state of a propagating optical field (i.e., a `flying
photon'). At node $B$ the pulse $\Omega _{B}^{in}(t)$ is applied to map the
state of the flying photon into an atom within the cavity, thereby realizing
the transfer of the state $\left\vert \protect\psi \right\rangle $ from $A$
to $B$ \protect\cite{boozer07}. (d) Distribution of entanglement using
ensembles of a large number of atoms \protect\cite{duan01}. A single-photon
pulse at node $A$ is coherently split into two entangled components that
propagate to nodes $B,C$ and are there coherently mapped by control fields $%
\Omega _{B,C}^{(in)}(t)$ into an entangled state between ensembles at $B,C$.
At later times, components of the entangled state can be retrieved from the
quantum memories by separate control fields $\Omega _{B,C}^{(out)}(t)$
\protect\cite{choi08}.}
\label{qnet}
\end{figure}

Within the broad context of QIS, quantum networks play an important role,
both for the formal analysis and the physical implementation of quantum
computation, communication, and metrology \cite%
{zoller05,bennett92,briegel00,spiller06,giovannetti04}. Figure \ref{qnet}(a)
illustrates a notional \textit{quantum network} based upon the original
proposals in Refs. \cite{briegel00,cirac97}. To create a quantum network,
quantum information is generated, processed, and stored locally in \textit{%
quantum nodes}. These nodes are linked by \textit{quantum channels} that
transport quantum states from site to site with high fidelity and that
distribute entanglement across the entire network. With relatively modest
processing capabilities, one could even envision a `quantum internet' that
could accomplish tasks that are otherwise impossible within the realm of
classical physics, including the distribution of `quantum software' \cite%
{preskill99,gottesman99}.

Apart from any particular algorithm, there is an important advantage to be
gained from \textit{quantum} as opposed to \textit{classical} connectivity
between nodes \cite{cirac99}. A network of quantum nodes linked by classical
channels comprised of $k$ nodes each with $n$ qubits has a state space of
dimension $k2^{n}$, whereas a fully quantum network linked by quantum
channels has an exponentially larger dimension $2^{kn}$. Quantum
connectivity also provides a potentially powerful means to overcome
size-scaling and error-correlation problems that would otherwise limit the
size of machines for quantum processing \cite{copsey03}. At any stage in the
development of quantum technologies, there will be a largest size attainable
for individual quantum processing units, which can be superseded by linking
such units together into a fully quantum network.

A quite different perspective of a quantum network is to view the nodes as
components of a physical system that interact by way of the quantum
channels. In this case, the underlying physical processes used for quantum
network protocols are adapted to simulate the evolution of quantum many-body
systems \cite{qmb}. For example, atoms localized at separate nodes can have
effective \textquotedblleft spin -- spin\textquotedblright\ interactions
catalyzed by single-photon pulses that travel along the channels between the
nodes \cite{duan05}. The \textquotedblleft quantum wiring\textquotedblright\
of the network allows a wide range for the effective Hamiltonian and
topology of the resulting \textquotedblleft lattice\textquotedblright .\
Moreover, the extension of entanglement across quantum networks can be
related to the classical problem of percolation \cite{acin07}.

These exciting opportunities motivate an examination of research related to
the actual physical processes for translating the abstract illustration in
Fig. \ref{qnet}(a) into reality. From diverse activities worldwide, I will
narrow the focus to provide an overview of current efforts and prospects for
harnessing optical processes at the level of single photons and atoms for
the reliable transport of quantum states across complex quantum networks. In
this context, we require that quantum states of a material system be mapped
to and from propagating optical fields as illustrated in Fig. \ref{qnet}(b),
thereby achieving quantum connectivity for the network, including
entanglement to enable quantum teleportation between nodes.

Such considerations are timely since scientific capabilities are now passing
the threshold from a learning phase with individual systems over the past
fifteen years and are advancing into a domain of rudimentary functionality
for quantum nodes connected by quantum channels. From a broad spectrum of
remarkable achievements, two important areas on which I will focus are
strong coupling of single photons and atoms within the setting of cavity
quantum electrodynamics (QED) \cite{miller05} and quantum information
processing with atomic ensembles \cite{duan01}. Critical to both these
examples are long-lived quantum memories provided by the atomic system and
efficient light-matter interfaces.

I apologize at the outset for my sins of omission related to alternative
approaches. In these early days, it is impossible to foresee which of many
exciting possibilities might lead to rudimentary networks with nontrivial
scientific capabilities much less to functional systems of technological
significance. Many physical systems are being investigated as surveyed in
Ref. \cite{zoller05} and described at the following websites: \textit{%
http://qist.lanl.gov/qcomp\_map.shtml}, \textit{%
http://www.scala-ip.org/public/}, and \textit{%
http://www.qubitapplications.com/}. However, my purpose is not to present an
exhaustive set of detailed descriptions, but rather to convey basic
principles for physical implementations of quantum networks by way of a few
examples in Quantum Optics. I strive thereby to stimulate the involvement of
a larger community in this endeavor, including for systems-level studies.

\section{A quantum interface between light and matter}

The principal scientific challenge in the quest to distribute quantum states
across a quantum network is the attainment of coherent control over\ the
interactions of light and matter at the single-photon level. Whereas atoms
and electrons have relatively large long-range interactions for their spin
and charge degrees of freedom, individual photons have interaction
cross-sections that are typically orders of magnitude too small for
nontrivial dynamics when coupled to single degrees of freedom for a material
system.

The endeavor to address this issue began in the optical physics community in
the 1990s with the development of theoretical protocols for the coherent
transfer of quantum states between atoms and photons within the setting of
cavity quantum electrodynamics (QED) \cite{cirac97,parkins93,vanenk98}. Other important
advances have been made in the past decade, including with atomic ensembles \cite{duan01,lukin-review,fleischhauer-review}. The reversible
mapping of quantum states between light and matter provides the basis for
quantum-optical interconnects and is a fundamental primitive for building
quantum networks. Although the original schemes for such interconnects are
fragile to experimental imperfections, a complete set of theoretical
protocols have subsequently been developed for the robust distribution of
quantum information over quantum networks, including, quite significantly,
the invention of the quantum repeater \cite{briegel00,briegel98} and
scalable quantum networks with atomic ensembles \cite{duan01}.

Fig. \ref{qnet}(b) illustrates a generic quantum interface between light and
matter described by the interaction Hamiltonian $H_{int}(t)$, where for
typical states $\left\langle H_{int}(t)\right\rangle \sim \hbar \chi (t)$,
with $\chi (t)$ the time-dependent coupling strength between the internal
material system and the electromagnetic field. Desiderata for a quantum
interface include that $\chi (t)$ should be \textquotedblleft
user-controlled\textquotedblright\ for clocking states to and from the
quantum memory (e.g., by way of an auxiliary laser), that the physical
processes employed should be robust in the face of imperfections (e.g., by
adiabatic transfer), and that mistakes should be efficiently detected and
fixed (e.g., with quantum error correction). In qualitative terms, the rate $%
\kappa $ which characterizes the bandwidth of the input-output channel
should be large compared to the rate $\gamma $ for any parasitic losses, and
both these rates should be small compared to the rate of coherent coupling, $%
\chi \gg \kappa \gg \gamma $.

Two specific examples of physical systems to realize a quantum interface and
distribute coherence and entanglement between nodes are shown in Fig. \ref%
{qnet}(c), (d). In the case of Fig. \ref{qnet}(c), single atoms are trapped
within optical cavities at nodes $A,B$ which are linked by an optical fiber.
External fields control the transfer of the quantum state $\left\vert \psi
\right\rangle $ stored in the atom at node $A$ to the atom at node $B$ by
way of photons that propagate from $A$ to $B$ \cite{cirac97,boozer07}. In
Fig. \ref{qnet}(d), a single photon pulse generated at node $A$ is
coherently split into two components and propagates to nodes $B,C$, where
the entangled photon state is coherently mapped into an entangled state
between collective excitations at the two nodes \cite{duan01,lukin-review,fleischhauer-review,choi08}.
Subsequent readout of entanglement from either or both of the memories at $%
B,C$ to photon pulses is implemented at the push of a button.

In the sections that follow, I will elaborate in somewhat more detail the
underlying physical processes for the implementation of quantum-optical
interconnects between matter and light. Particular attention will be given
to the two examples in Fig. \ref{qnet}(c), (d).

\section{Cavity quantum electrodynamics}

At the forefront of the endeavor to achieve strong, coherent interactions between light and matter has been the area of cavity quantum
electrodynamics (QED) \cite{berman94}. In both the optical \cite{miller05,wilk07} and microwave \cite{meschede85,meystre92,walther04,haroche05,guerlin07}
domains, strong coupling of single atoms and photons has been achieved by utilizing electromagnetic resonators of small mode volume $V_{m}$ with
quality factors $Q\sim 10^{7}-10^{11}$. Extensions of cavity QED to other systems \cite{vahala-review} include quantum dots coupled to
micropillars and photonic bandgap cavities \cite{khitrova06,barclay07,shields07,englund07b} and Cooper-pairs interacting with superconducting resonators (i.e., `circuit' QED
reviewed in Ref. \cite{schoelkopf08}).

\begin{figure}[h]
\begin{center}
\includegraphics[width=13cm]{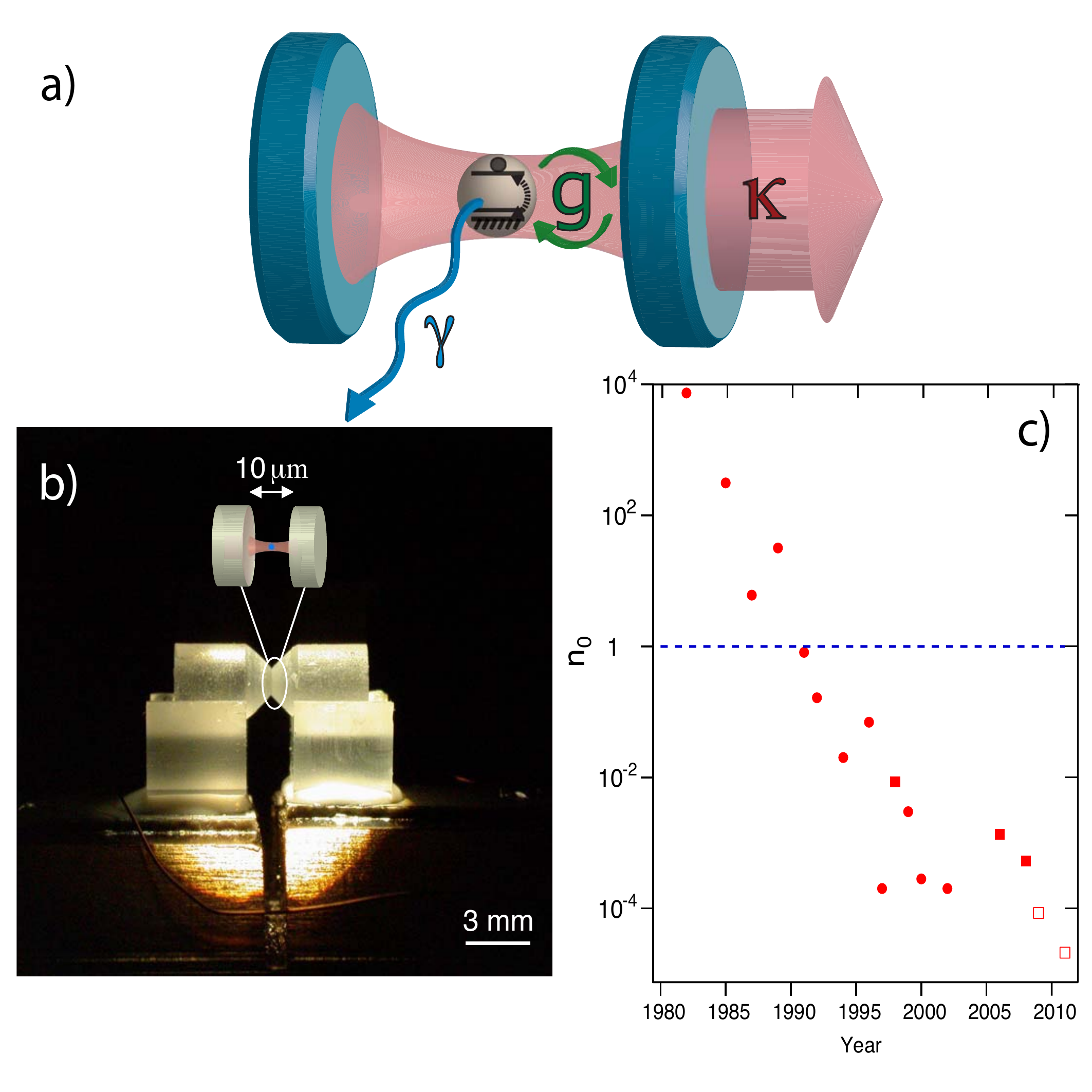}
\end{center}
\caption{Elements of cavity QED. (a) A simple schematic of an atom-cavity
system depicting the three governing rates $(g,\protect\kappa ,\protect%
\gamma )$ in cavity QED, where $g\sim \protect\chi $ from Fig. \protect\ref%
{qnet}. (b) Photograph of two mirror substrates that form a Fabry-Perot
cavity shown schematically above. The cavity has length $l=10$ $\protect\mu $%
m, waist $w_{0}=12$ $\protect\mu $m transverse to the cavity axis, and
finesse $F\simeq 5\times 10^{5}$. The supporting structure serves for active
servo control of the cavity length to $\protect\delta l\simeq 10^{-14}$ m
\protect\cite{miller05}. (c) Critical photon number $n_{0}$ versus time for
a series of experiments in cavity QED, first at the Unversity of Texas at Austin and then in the Caltech Quantum Optics Group.
These experiments involve both spherical mirror Fabry-Perot cavities
(circles) and the whispering gallery modes of monolithic $SiO_{2}$
resonators (squares). The points in $2006,2008$ are from Refs. \protect\cite%
{aoki06,dayan08} demonstrated for a $SiO_{2}$ microtoroidal resonator, while
those in $2009,2011$ are projections for a microtoroid \protect\cite%
{spillane05}.}
\label{cQED}
\end{figure}

Figure \ref{cQED}(a, b) depicts a single atom located inside an optical
resonator, for which strong coupling of atom and photon requires that a
single intracavity photon creates a \textquotedblleft
large\textquotedblright\ electric field. Stated more quantitatively, if the
coupling frequency of one atom to a single mode of an optical resonator is $%
g$ (i.e., $2g$ is the one-photon Rabi frequency), then

\begin{equation}
g=\sqrt{\frac{|\vec{\varepsilon}\cdot \vec{\mu}_{0}|^{2}\omega _{C}}{%
2\hbar \epsilon _{0}V_{m}}}\text{.}  \label{g0}
\end{equation}%
Here, $\vec{\mu}_{0}$ is the transition-dipole moment between the relevant
atomic states with transition frequency $\omega _{A}$, and $\omega
_{C}\simeq \omega _{A}$ is the resonant frequency of the cavity field with
polarization vector $\vec{\varepsilon}$ and mode volume $V_{m}$. Experiments
in cavity QED explore strong coupling with $g\gg (\gamma ,\kappa )$,
where $\gamma $ is the atomic decay rate to modes other than the cavity mode
and $\kappa $ is the decay rate of the cavity mode itself. Expressed in the
language of traditional optical physics, the number of photons required to
saturate the intracavity atom is $n_{0}\sim \gamma ^{2}/g^{2}$, and the
number of atoms required to have an appreciable effect on the intracavity
field is $N_{0}\sim {\kappa \gamma /g^{2}}$. Strong coupling in cavity
QED moves beyond traditional optical physics for which $(n_{0},N_{0})\gg 1$
to explore a qualitatively new regime with $(n_{0},N_{0})\ll 1$ \cite%
{miller05}.

A variety of approaches have been employed to achieve strong coupling in cavity QED over the past three decades, with reviews of much of the
early work available in Ref.\cite{berman94}. Of particular note in the development of this field were the pioneering experiments of H. Walther
\cite{meschede85}, who combined Rydberg atoms with superconducting cavities ($Q\sim 10^{10}$) to advance cavity QED into a regime of strong
coupling in the microwave domain \cite{walther04,haroche05}. A recent spectacular achievement is the observation of the step-by-step collapse of
the quantum state of an intracavity microwave field \cite{guerlin07}.

In the optical domain, a route to strong coupling is the use of high-finesse
optical resonators ($F\sim 10^{5}-10^{6}$) and atomic transitions with large
$\vec{\mu}_{0}$ (i.e., oscillator strengths near unity). Figure \ref{cQED}%
(c) provides an illustration of progress along this avenue, with research
now well into the domain $(n_{0},N_{0})\ll 1$.

Of course as the cavity volume $V_{m}$ is reduced to increase $g$ as in Eq. \ref{g0}, the requirement for atomic localization necessarily
becomes more stringent. Not surprisingly, central to activities in cavity QED over the past decade have been efforts to trap and localize atoms
within high-finesse optical cavities in a regime of strong coupling, with the initial demonstration in 1999 \cite{ye99b}. Subsequent advances
are reviewed
in Ref. \cite{ye08}, including trap lifetimes $\gtrsim 10$ s \cite%
{hijlkema07,fortier07}. Quantum control has now been achieved over both
internal (i.e., atomic dipole and cavity field) and external (i.e., atomic
motion) degrees of freedom for a strongly coupled atom-cavity system \cite%
{boozer06}. An exciting prospect is cavity QED with single trapped ions,
where the boundary for strong coupling has been reached \cite{keller04}.

\subsection{Coherence and entanglement in cavity QED}

An application of these capabilities\ to quantum networks is the generation
of single photons \textquotedblleft on demand\textquotedblright\ (see Box
1). By way of strong coupling of the cavity field to an atomic transition,
an external control field $\Omega (t)$ transfers one photon into the cavity
mode and thence to free-space by way of the cavity output mirror, leading to
a single-photon pulse $\left\vert \phi _{1}(t)\right\rangle $ as a
collimated beam. The temporal structure (amplitude and phase) of the
resulting \textquotedblleft flying photon\textquotedblright\ $\left\vert
\phi _{1}(t)\right\rangle $\ can be tailored by way of the control field $%
\Omega (t)$ \cite{cirac97,kuhn99,duan04}, with the spatial structure of the
wavepacket set by the cavity mode $\psi (\vec{r})$.

Several experiments have confirmed the essential aspects of this process for
the deterministic generation of single photons \cite%
{hijlkema07,keller04,mckeever04}. Significantly, in the ideal, adiabatic
limit, the atomic excited state $\left\vert e\right\rangle $ is not
populated \cite{bergmann98}. By deterministically generating a bit stream of
single-photon pulses from single, trapped atoms, these experiments provide a
first step in the evolution of quantum networks based upon \textit{flying}
photons.

Relative to single-photon generation by diverse other systems \cite{lounis05}%
, a distinguishing aspect of the dark-state protocol discussed in Box 1 is
that it should be \textit{reversible}, so that a photon emitted from one
system \textit{A} can be efficiently transferred to another system \textit{B}
by simply applying the time-reversed (and suitably delayed) field $\Omega
(t) $ at \textit{B}, as illustrated in Figure \ref{qnet}(c).

Such an advance was reported in Ref. \cite{boozer07} by implementing the
reversible mapping of a coherent optical field to and from internal states
of a single, trapped Cesium atom. As described by Eq. \ref{in-out} in Box 1,
the incident field $\left\vert \phi _{field}(t)\right\rangle \sim
c_{0}\left\vert 0\right\rangle _{field}+c_{1}\left\vert 1\right\rangle
_{field}$ was approximated by a coherent state $\lambda (t)$ with $\bar{n}%
=1.1$ photons and was transferred into a coherent superposition of atomic
states. The stored atomic state was then coherently mapped back into a
propagating field $\beta (t)$. Although there were imperfections in this
experiment \cite{boozer07}, it provides the initial verification of the
fundamental primitive upon which the protocol in Ref. \cite{cirac97} is
based.

The discussion in Box 1 as well as related possibilities \cite{duan05,duan04}
rely upon strong coupling between an atom and a \textit{single} polarization
of the intracavity field. However, by extending these ideas to the two
polarization eigenmodes of the cavity for a given TEM$_{00}$ longitudinal
mode, it is possible to generate entanglement between internal atomic states
and the polarization state of a coherently generated photon \cite%
{lange00,duan03b,sun04}. An initial control field $\Omega _{1}(t_{1})$
results in an entangled state between internal states of the atom $%
\left\vert b_{\pm }\right\rangle $ and the polarization state of a flying
photon\ $\left\vert \phi _{field}^{\pm }(t_{1})\right\rangle $ coherently
generated by the coupled atom-cavity system. Application of a second control
field $\Omega _{2}(t_{2})$ returns the atom to its initial (unentangled)
state while generating a second flying photon $\left\vert \xi _{field}^{\pm
}(t_{2})\right\rangle $, thereby leading to entanglement between the
polarizations $\sigma _{\pm }$ of the two fields $\phi _{field}^{\pm },\xi
_{field}^{\pm }$ emitted at times $t_{1},t_{2}$.

Precisely such a sequence of operations $\Omega _{1,2}(t_{1,2})$ has been
applied to single Rubidium atoms falling through a high-finesse optical
cavity \cite{wilk07}. In an experimental \textit{tour de force}, these
authors were able to create entangled photons with time separation $\tau
=t_{2}-t_{1}$ limited by the atomic transit time. Although the atoms arrived
randomly, the protocol itself is intrinsically deterministic and represents
a significant advance toward the generation and distribution of entangled
states for quantum networking. With trapped atoms, it will be possible to
generate entangled states at user selected times $(t_{1},t_{2})$ at the
\textquotedblleft push of button.\textquotedblright\ Moreover, the scheme is
inherently reversible, so that the entangled state between atom and field
can be used to distribute entanglement to a second atom-cavity system in a
network.

In a broader context, important advances have been made in the generation
and transfer of quantum states with other physical systems. For example,
single photons generated by a quantum dot coupled to a photonic-bandgap
cavity have been transmitted to a second, `target' cavity via an on-chip
waveguide \cite{englund07}. Ref. \cite{schoelkopf08} reviews the spectacular
progress made in wiring quantum circuits by way of superconducting microwave
cavities and coherent transmission along a quantum bus.

This discussion provides a preview of coming experimental advances towards
the realization of quantum networks based upon interactions in cavity QED.
Diverse theoretical protocols have been developed but have awaited the
maturation of experimental capabilities, which we are now witnessing.
Included are the sequential generation of entangled multiqubit states \cite%
{schon05}, the teleportation of atomic states from one node to another \cite%
{vanenk98}, photonic quantum computation by photon-photon interactions at
the nodes \cite{duan04}, and reversible mapping of quantum states of atomic
motion to and from light \cite{parkins99}. Certainly, new technical
capabilities beyond conventional Fabry-Perot cavities will be required to
enable such scientific investigations, with Box 2 discussing some candidate
systems.

\section{Quantum networks with atomic ensembles}

An area of considerable activity in the quest to distribute coherence and
entanglement across quantum networks has been the interaction of light with
atomic ensembles comprised of a large collection of identical atoms. For the
regime of continuous variables, entanglement has been achieved between two
atomic ensembles each consisting of $\sim 10^{12}$ atoms \cite{julsgaard01},
and quantum teleportation of light to matter demonstrated for the mapping of
coherent optical states to the collective spin states of an atomic memory
\cite{sherson06}. More general activities in this area are covered in Ref.
\cite{polzik-book}.

Here, I will focus instead on the regime of discrete variables with photons
and atomic excitations taken one by one. Research in this area is based upon
the remarkable theoretical protocol in Ref. \cite{duan01} (hereafter
referred to as `DLCZ'), which presented a realistic scheme for entanglement
distribution by way of a quantum-repeater architecture \cite%
{briegel00,briegel98}.

Fundamental to the DLCZ protocol is the generation and retrieval of single
`spin' excitations within an ensemble of a large number of atoms (see Box 3)
\cite{raymer85}. In concert with photoelectric detection, a first laser
pulse creates a single excitation $\left\vert 1_{a}\right\rangle $ stored
collectively within the atomic ensemble. At a later time, a second pulse
deterministically converts excitation stored within the atomic memory in the
state $\left\vert 1_{a}\right\rangle $ into a propagating field, denoted as
field $2$.

\begin{figure}[h]
\begin{center}
\includegraphics[width=11cm]{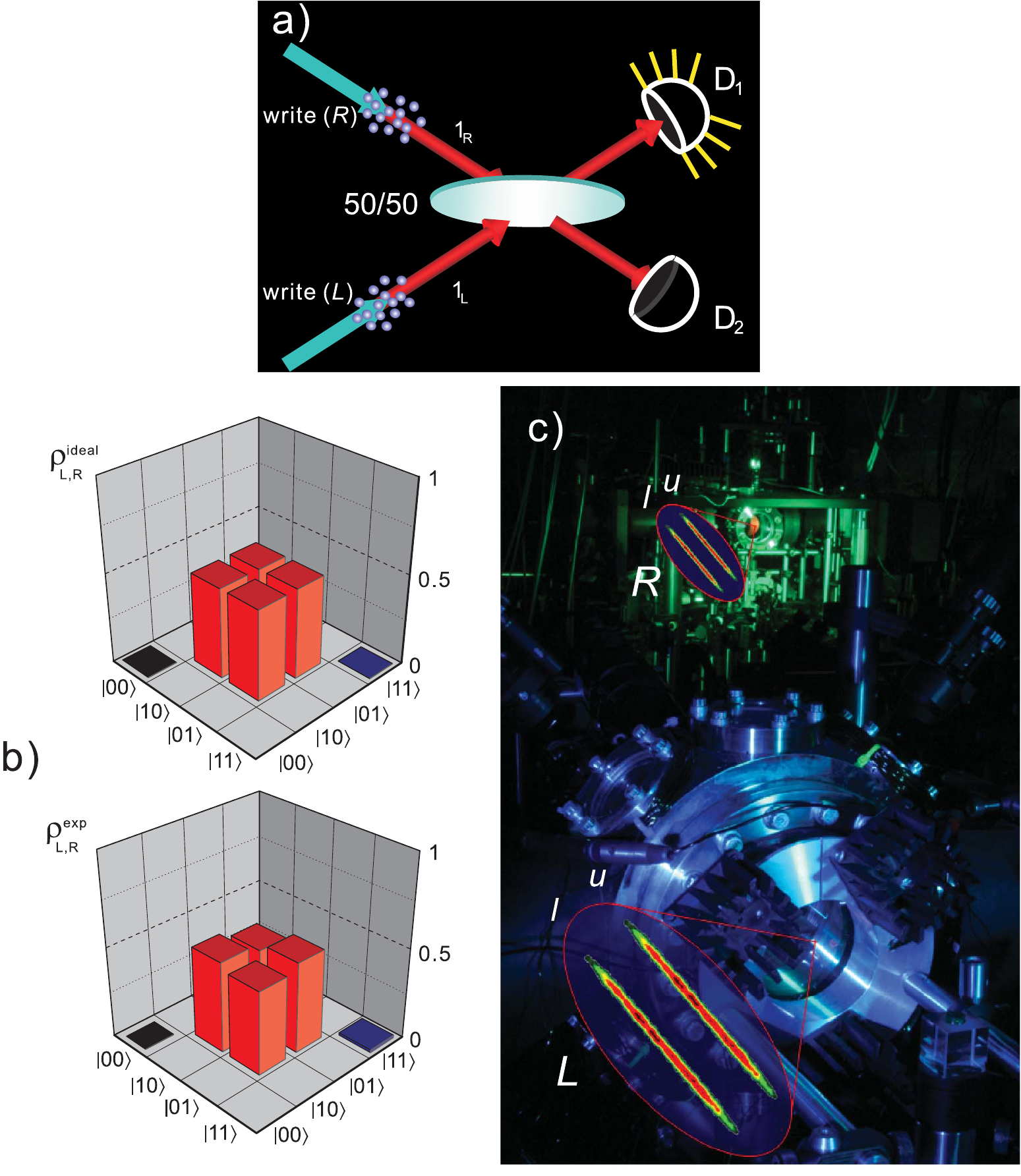}
\end{center}
\caption{Fundamentals of the DLCZ protocol \protect\cite{duan01}. a) Measurement-induced entanglement between two atomic ensembles $L,R$
\protect\cite{duan01,chou05}. Synchronized laser pulses incident on the ensembles (write beams $L,R$) generate small amplitudes for spontaneous
Raman scattering \protect\cite{raymer85}, denoted as $1_{L},1_{R} $. These fields interfere at a 50\%-50\% beam splitter, with outputs directed
to single-photon detectors $D_{1},D_{2}$. A measurement event at
either detector projects the $L,R$ ensembles into the entangled state $%
\left\vert \Psi _{L,R}\right\rangle $ with one quantum of excitation shared
remotely between the ensembles. Entanglement is stored in the quantum memory
provided by the ensembles and can be subsequently converted to propagating
light pulses by a set of read laser pulses (Box 3). b) Experimentally
determined elements of the density matrix $\protect\rho _{L,R}^{\exp }$ for
entanglement between two atomic ensembles \protect\cite{laurat07b},
corresponding to concurrence $C=0.9\pm 0.3$, where $C=0$ for an unentangled
state. For comparison the density matrix $\protect\rho _{L,R}^{th}$ for the
ideal state $\left\vert \Psi _{L,R}\right\rangle $ is shown with concurrence
$C=1$. c) Photograph of the laboratory setup for entanglement of two pairs
of atomic ensembles for the realization of functional quantum nodes $L,R$
separated by $3$ m \protect\cite{chou07}. The four elongated ovals each show a cylinder of $10^{5}$\ Cesium atoms which forms an atomic ensemble at
each site. Entangled states $\left\vert \Psi _{L,R}^{u}\right\rangle
,\left\vert \Psi _{L,R}^{l}\right\rangle $ between the upper $u$ and lower $l
$ pairs at the $L,R$ nodes are generated and stored in an asynchronous
fashion for each pair $u,l$ as in a). Atomic excitations for the pairs $%
L_{u},L_{l}$ and $R_{u},R_{l}$ are subsequently converted to flying photons
at each node with polarization encoding that leads to violation of a Bell
inequality \protect\cite{chou07}. The entire experiment functions under the
quantum control of single detection events.}
\label{entanglement}
\end{figure}

The basic processes illustrated in Box 3 can be extended to create an
entangled pair of ensembles as in Fig.~\ref{entanglement}(a) \cite{duan01}.
A pair of ensembles $L,R$ is illuminated by separate write pulses, leading
to the overall state $\left\vert \phi _{L,R}\right\rangle =\left\vert \phi
_{a,1}^{L}\right\rangle \otimes \left\vert \phi _{a,1}^{R}\right\rangle $,
where $\left\vert \phi _{a,1}^{L}\right\rangle ,\left\vert \phi
_{a,1}^{R}\right\rangle $ are entangled states between atomic excitation and
photon number for field $1$ for the $L,R$ ensembles (see Eq. \ref{phia1} in
Box 3). The scattered fields $1_{L},1_{R}$ from the two ensembles are
combined on a 50-50 beamsplitter, with outputs directed to two
photodetectors $D_{1},D_{2}$. In the ideal case and to lowest order in $p$,
a photoelectric detection event at either detector projects the ensembles
into the entangled state%
\begin{equation}
\left\vert \Psi _{L,R}\right\rangle =\frac{1}{\sqrt{2}}[\left\vert
0_{a}\right\rangle _{L}\left\vert 1_{a}\right\rangle _{R}\pm e^{i\eta
_{1}}\left\vert 1_{a}\right\rangle _{L}\left\vert 0_{a}\right\rangle _{R}]%
\text{,}  \label{LRent}
\end{equation}%
with the sign $+$ or $-$ set by whether $D_{1}$ or $D_{2}$ records the
event. The phase $\eta _{1}$ is determined by the difference of phase shifts
along the two channels, $\eta _{1}=\beta _{L}-\beta _{R}$ \cite{chou05},
which must be stable.

The state $\left\vert \Psi _{L,R}\right\rangle $ is generated in a
probabilistic but heralded fashion from quantum interference in the
measurement process \cite{dicke81,cabrillo99,bose99}. That is, detection of a photon from one or the other
atomic ensemble in an indistinguishable fashion results in an entangled
state with one collective spin excitation shared coherently between the
ensembles. Necessarily, because $p\ll 1$, any given trial with the write
pulses is unlikely to produce a detection event at $D_{1},D_{2}$, with such
failed trials requiring the system to be reinitialized. However, a
photoelectric detection event at $D_{1},D_{2}$ unambiguously heralds the
creation of the entangled state $\left\vert \Psi _{L,R}\right\rangle $.
Limited by the coherence time between the levels $\{\left\vert
g\right\rangle _{i},\left\vert s\right\rangle _{i}\}$ \cite{laurat07b}, this
entangled state is stored in the quantum memory provided by the $L,R$
ensembles and available `on demand'\ for subsequent tasks, such as
entanglement connection \cite{duan01,laurat07a}.

Although the preceding discussion is for an ideal case and neglects
higher-order terms, the DLCZ protocol is designed to be resilient to
important sources of imperfections, including losses in propagation and
detection, as well as detector dark counts. Indeed, the scheme functions
with \textquotedblleft built-in entanglement purification\textquotedblright\
\cite{duan01} and enables entanglement to be extended beyond the separation
of two ensembles in an efficient and scalable fashion. Theoretical
extensions \cite{jiang07,zhao07,sangouard08} of the DLCZ protocol have examined
related network architectures for optimizing scalability in view of actual
laboratory capabilities, to which I next turn.

\subsection{Coherence and entanglement with atomic ensembles}

The publication of the DLCZ protocol in 2001 led to the rapid development of
a worldwide community with by now significant achievements in the creation
and distribution of entanglement. The initial, enabling steps in the
implementation of the protocol of DLCZ were the observations of quantum
correlations both for single photon pairs \cite{kuzmich03,balic05} and for
large photon number ($10^{3}-10^{4}$) \cite{vanderwal03} generated in the
collective emission from atomic ensembles. Single photons were generated by
the efficient mapping of stored, collective atomic excitation to propagating
wavepackets for field $2$, as in Box 3 \cite%
{chou04,laurat06,thompson06,kuzmich06,chen06}. Conditional readout
efficiencies of $74\%$ in free space \cite{du08} and $84\%$ \cite%
{simon07} in a cavity were realized for the state transfer from a single,
collective \textquotedblleft spin\textquotedblright\ excitation stored in
the atomic ensemble to a single photon for field $2$.

With these capabilities for coherent control of collective atomic emission,
heralded entanglement between distant ensembles was achieved in 2005 \cite%
{chou05}, with the entangled state stored in quantum memories (i.e., the two
ensembles) located in distinct apparatuses separated by $3$ m. After a
programmable delay, entanglement was confirmed by mapping the state of the
atoms to optical fields and by measuring mutual coherence and photon
statistics for these fields. Although the degree of entanglement between the
ensembles was small in the initial experiment \cite{chou05}, more recent
work \cite{laurat07b} has led to the inference $C=0.9\pm 0.3$ for the
concurrence $C$ \cite{wootters98} of entanglement stored between $L,R$
ensembles, with the associated density matrix $\rho _{L,R}$\ shown in Fig. %
\ref{entanglement}(b).

The DLCZ protocol is based upon a quantum repeater architecture involving
independent operations on parallel chains of quantum systems \cite{duan01},
with scalability relying critically upon conditional control of quantum
states stored in remote quantum memories \cite{felinto06}. The experiment
shown in Fig. \ref{entanglement}(c) \cite{chou07} took an important step
towards this goal by achieving the minimal functionality required for
scalable quantum networks. Heralded entanglement was created asynchronously
between a qubit located at node $L$\ and one at node $R$, where each qubit
was encoded by a pair of upper $u$ and lower $l$ atomic ensembles. The
entangled states at the $(L,R)$\ nodes were stored in the quantum memory
provided by the ensembles, and then efficiently mapped to propagating light
fields resulting in an effective polarization entangled state, as verified
by the measured violation of a Bell inequality for these light fields. This
setup was also used for the initial investigation of entanglement swapping
\cite{laurat07a} to demonstrate coherence between ensembles that had no
direct interaction.

Beyond DLCZ's proposal for measurement-induced entanglement, it is also
possible to achieve deterministic mapping of quantum states of light to and
from atomic ensembles by way of electromagnetically induced transparency
(EIT) \cite{lukin-review,fleischhauer-review,harris97}. The pioneering work in Refs. \cite%
{Hau01,phillips01} demonstrated the storage and retrieval of classical pulses to and from an ensemble. Refs. \cite{Chaneliere05,Eisaman05}
extended this work into the quantum regime of single photons. Entanglement between two ensembles coupled to a cavity mode was achieved in Ref.
\cite{vuletic07} by adiabatic transfer of excitation, thereby providing a means for on-demand entanglement. To assist the distribution of
entanglement over quantum networks, Ref. \cite{choi08} realized the reversible mapping of photonic entanglement into and out of pairs of quantum
memories by an EIT\ process illustrated in Fig. \ref{qnet}(d).

Contemporary with these experiments which achieved heralded \cite%
{duan01,chou05,laurat07b,chou07} and deterministic \cite{choi08,vuletic07} entanglement have been a variety of experiments based upon
entanglement as a post diction \cite{vanenk05}, for which a physical state is not available for utilization in a scalable network but which are
nonetheless significant. An important advance in this regard is the work in Ref. \cite{chen08}, which
employed a pair of ensembles for entanglement generation to achieve \textit{%
a posteriori} teleportation of light to an atomic memory.

There has also been considerable effort devoted to the detailed
characterization of decoherence for stored atomic excitation and
entanglement \cite{chou07,laurat07b,chen08}. Decoherence of entanglement
between distinct atomic ensembles was observed in the decay of the violation
of a Bell inequality \cite{chou07} and of the fidelity for teleportation
\cite{chen08}. By way of measurements of concurrence $C(t)$, Ref. \cite%
{laurat07b} presented quantitative characterizations for the relationship of
the global evolution of the entangled state to the temporal dynamics of
various local correlations.

\section{Extending entanglement for quantum networks}

The entangled states created so far both in cavity QED\ and by way of the
DLCZ\ protocol are between pairs of systems, so-called bipartite
entanglement, for which definitive procedures exist for operational
verification. Certainly the creation of more general classes of entangled
states shared among $N>2$ systems would be of great interest. In this
regard, an important area of ongoing research is the development of
theoretical protocols amenable to laboratory implementation for entanglement
verification of quantum states distributed over the nodes of a quantum
network. However, as we progress toward more complex quantum networks, the
issue of entanglement verification becomes increasingly problematic.
Theoretical tools and experimental capabilities do not at present exist for
characterizing the general states of quantum networks.

Perhaps surprisingly, a nontrivial task will be simply to answer the
question \textquotedblleft Does it work?\textquotedblright \textit{\ }That
is, as even moderately complex quantum networks are realized in the
laboratory, it will become ever more difficult to assess quantitatively the
characteristics of the network, including such basic questions as
\textquotedblleft Does entanglement extend across the whole
network?\textquotedblright\ On the one hand, we might follow a strategy
motivated by the underlying physical processes upon which the network is
based and attempt to determine, for example, the density matrix $\rho (t)$
for the network. This course of action surely fails because of the
exponential growth in $\rho (t)$ with the size of the network.

We might instead follow a strategy based upon more functional issues of
algorithmic capability. That is, we could attempt to implement a quantum
algorithm for computation or communication to test whether the purported
quantum network actually accomplishes anything beyond the capabilities of
any classical counterpart. This course is problematic since the advantage of
a quantum network may only be realized above some threshold in the size of
the network. Furthermore, from an experimental perspective, this strategy
does not offer much in the way of diagnostics for `fixing' the network when
it fails.

A less obvious approach might be to adopt more seriously the perspective of
a quantum network as a quantum many body system and to search for more
`physical' characteristics of the network (e.g., the scaling behavior of
pair correlation functions and multipartite entanglement). Indeed, an active
area of research is the nature of entanglement for systems that undergo
quantum phase transitions, including the pioneering advance for 1-D spin
chains in Ref. \cite{vidal03}.

\section{Conclusion}

Although important progress has been made, there remain many scientific
challenges that require new theoretical\ insights and experimental
capabilities in the quest to realize functional quantum networks. Certainly,
the current state of the art is primitive relative to that required for the
robust and scalable implementation of sophisticated network protocols,
whether over short or long distances. The realization of quantum memories,
local quantum processing, quantum repeaters, and error-corrected
teleportation are very ambitious goals. There is nevertheless considerable
activity worldwide directed towards these important goals.

While my discussion has considered separately cavity QED-based networks\ and
those implemented via the DLCZ protocol, clearly quantum networks will
evolve as heterogeneous entities. For example, the same protocol that
creates the entangled state $\left\vert \Psi _{L,R}\right\rangle $ between
two ensembles in Eq. \ref{LRent} can be employed to create the entangled
state%
\begin{equation}
\left\vert \Pi _{A,E}\right\rangle \sim \lbrack \left\vert
1_{A}\right\rangle \left\vert 0_{E}\right\rangle \pm \left\vert
0_{A}\right\rangle \left\vert 1_{E}\right\rangle ]\text{ ,}  \label{pice}
\end{equation}%
with one excitation shared jointly by an atom in a cavity $\left\vert
1_{A}\right\rangle $ and an atomic ensemble $\left\vert 1_{E}\right\rangle $%
. A critical task is the development of unambiguous procedures for
entanglement verification, which is a nontrivial undertaking that has not
always been carried out correctly \cite{vanenk05}.

I have used quantum networks as unifying theme but stress that the research
described has much broader import, including for advancing our understanding
of quantum dynamical systems and, for the particular cases considered,
creating new physics from controlled nonlinear interactions of single
photons and atoms. Altogether, these are exciting times in Quantum
Information Science as we pass from a regime of individual `Lego blocks'
(e.g., a single atom-cavity system) into the realm of complex quantum
systems assembled from many such units from the ground up. One important
area of this scientific enterprise is the realization of quantum networks. I
have every confidence that extending entanglement across quantum networks
will create wonderful scientific opportunities for the exploration of
physical systems that have not heretofore existed in the natural world.

\pagebreak

\bigskip

\textbf{Acknowledgement -- }I gratefully acknowledge the contributions made
by members of the Caltech Quantum Optics Group, and especially K. S. Choi,
B. Dayan, and R. Miller for their efforts on this manuscript. I am indebted
to J. P. Preskill and S. J. van Enk for their insights. This research is
supported by the National Science Foundation, by IARPA, and by Northrop
Grumman Space Technology.\bigskip

\textbf{Correspondence} -- Correspondence and requests for materials should
be addressed to HJK. (email: hjkimble@caltech.edu).

\pagebreak

\section{Box 1 - Mapping quantum states between atoms and photons}

Reversible state transfer between light and a single trapped atom can be
achieved by way of the mappings $\left\vert b\right\rangle \left\vert
1\right\rangle \rightarrow \left\vert a\right\rangle \left\vert
0\right\rangle $ and $\left\vert a\right\rangle \left\vert 0\right\rangle
\rightarrow \left\vert b\right\rangle \left\vert 1\right\rangle $ for the
coherent absorption and emission of single photons as illustrated in parts
(i) and (ii) of the figure \cite{boozer07}. Here $\left\vert a\right\rangle
,\left\vert b\right\rangle $ represent internal states of the atom with
long-lived coherence (e.g., atomic hyperfine states in the $6S_{1/2},F=3$
and $F=4$ manifolds of atomic Cesium) and $\left\vert 0\right\rangle
,\left\vert 1\right\rangle $ are Fock states of the intracavity field with $%
n=0,1$ excitations. The transition $\left\vert b\right\rangle
\leftrightarrow \left\vert e\right\rangle $ is strongly coupled to a mode of
an optical cavity with interaction energy $\hbar g$. In this simple
setting, the interaction Hamiltonian for atom and cavity field has a
\textquotedblleft dark state\textquotedblright\ (i.e., no excited state
component $\left\vert e\right\rangle $) \cite{bergmann98} given by \cite%
{parkins93}%
\begin{equation}
\left\vert D\right\rangle =\cos \theta \left\vert a\right\rangle \left\vert
0\right\rangle +\sin \theta \left\vert b\right\rangle \left\vert
1\right\rangle \text{,}  \tag{1}  \label{D}
\end{equation}%
where $\cos \theta =[1+\frac{\Omega ^{2}(t)}{g^{2}}]^{-1/2}$ with $%
\Omega (t)$ as a classical control field. For $\Omega (t=0)=0$, we have $%
\left\vert D\right\rangle =\left\vert a\right\rangle \left\vert
0\right\rangle $, while for $\Omega (t\rightarrow \infty )\gg g$, $%
\left\vert D\right\rangle \rightarrow \left\vert b\right\rangle \left\vert
1\right\rangle $.

In (i), by adiabatically ramping a control field $\Omega _{1}(t)\gg g$
from \textit{on} to \textit{off} over a time scale $\Delta t$ slow compared
to $1/g$, the atomic state is mapped from $\left\vert b\right\rangle $
to $\left\vert a\right\rangle $ with the accompanying coherent absorption of
$1$ intracavity photon. Conversely, in (ii) by turning a control field $%
\Omega _{2}(t)$ from \textit{off} to \textit{on}, the atomic state is mapped
from $\left\vert a\right\rangle $ to $\left\vert b\right\rangle $ with the
transfer of $1$ photon into the cavity mode.

As shown in (a)-(c) in the figure, these two processes can be combined to
achieve the coherent transfer of the state of a propagating optical field $%
\lambda (t)=\left\vert \phi _{field}(t)\right\rangle $ into and out of a
quantum memory formed by the atomic states $\left\vert a\right\rangle
,\left\vert b\right\rangle $ \cite{boozer07}. In the ideal case, the mapping
is specified by%
\begin{equation}
\left\vert \phi _{field}(t)\right\rangle \left\vert b\right\rangle
\rightarrow ^{(a)}\left\vert 0\right\rangle (c_{1}\left\vert a\right\rangle
+c_{0}\left\vert b\right\rangle )\text{ }\cdots ^{(b)}\cdots \rightarrow
^{(c)}\left\vert \phi _{field}(t+\tau )\right\rangle \left\vert
b\right\rangle \mathrm{.}  \tag{2}  \label{in-out}
\end{equation}%
where the field state is taken to be a coherent superposition of zero and
one photon, $\left\vert \phi _{field}(t)\right\rangle =\mathcal{E}%
(t)[c_{0}\left\vert 0\right\rangle _{field}+c_{1}\left\vert 1\right\rangle
_{field}]$. $\mathcal{E}(t)$ describes the envelope of the field external to
the cavity, with $\int |\mathcal{E}(t)|^{2}dt=1$. Given timing information
for incoming field $\left\vert \phi _{field}(t)\right\rangle $, step (a) in
this process is accomplished by adiabatically ramping the control field $%
\Omega _{1}(t)$ from \textit{on} to \textit{off}. In (b), the internal
states of the atom provide a long-lived quantum memory. At a user-selected
later time $t+\tau $, step (c) is initiated by turning $\Omega _{2}(t+\tau )$
from \textit{off} to \textit{on}, thereby coherently mapping the atomic
state $c_{1}\left\vert a\right\rangle +c_{0}\left\vert b\right\rangle $ back
to the \textquotedblleft flying\textquotedblright\ field state $\beta
(t)=\left\vert \phi _{field}(t+\tau )\right\rangle $.

\begin{center}
\includegraphics[width=13cm]{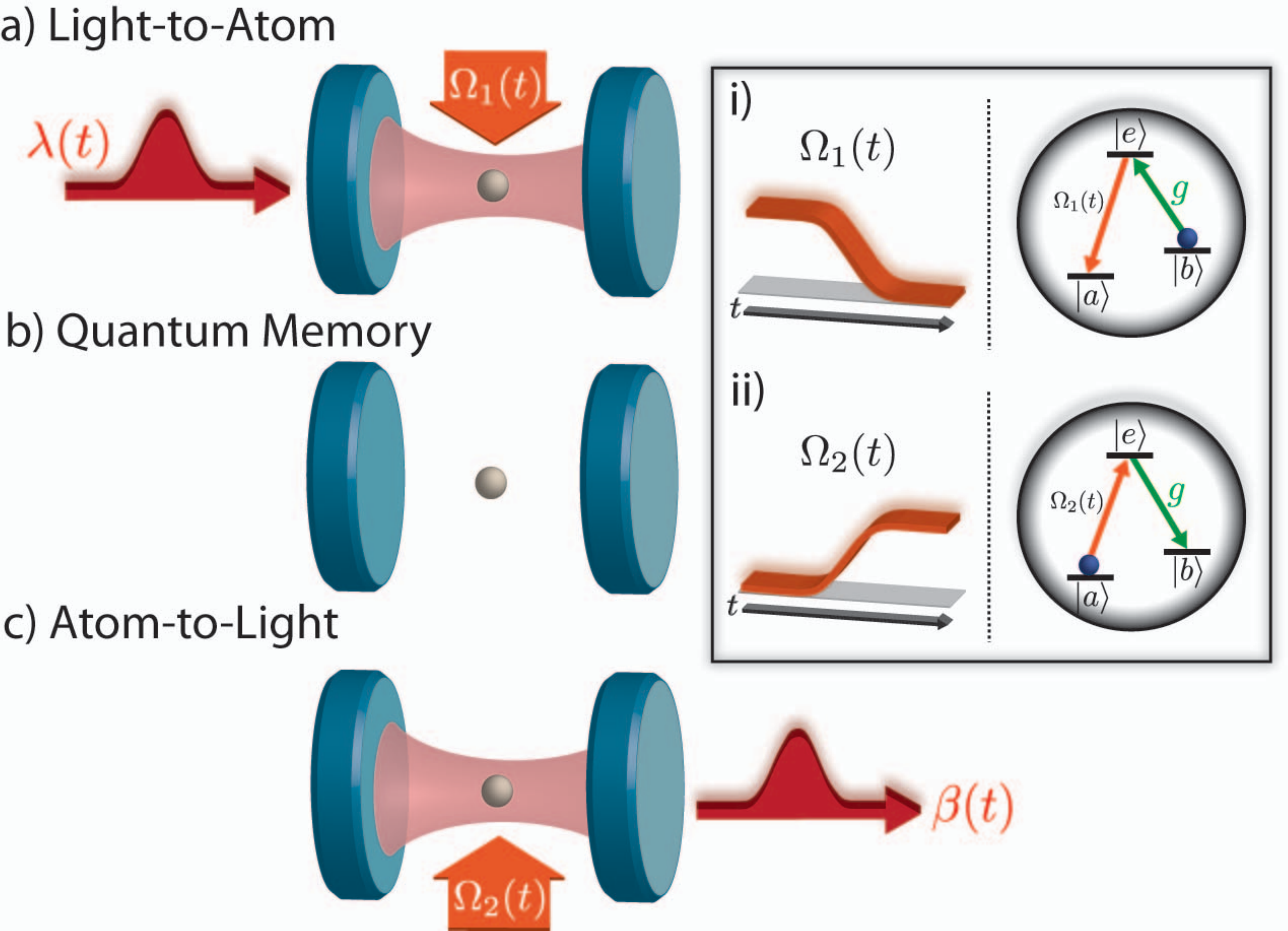}
\end{center}

\pagebreak

\section{Box 2 - A new paradigm for cavity QED}

The realization of large-scale quantum networks \cite{briegel00,cirac97} requires the capability to interconnect many quantum nodes\ over
quantum channels, for which conventional Fabry-Perot configurations are ill suited \cite{hood01}. There have correspondingly been efforts to develop of
alternative microcavity systems \cite{vernooy98,vahala-review}, both for single atoms \cite%
{aoki06,dayan08,trupke07,treutlein06} and atom-like systems \cite{xu07,park06}, such as nitrogen vacancy
centers in diamond \cite{park06}. A quantitative comparison of many
candidate systems is provided by Table I in Ref. \cite{spillane05}.

A remarkable resonator for this purpose is the microtoroidal cavity formed
from fused silica (SiO$_{2}$) shown in the figure \cite{armani03,spillane03}%
. Such a resonator supports a whispering gallery mode (WGM) \cite%
{braginsky89} circulating around the outer circumference of the toroid with
an evanescent field external to the resonator, as shown in part (c) of the
figure. Because of the small mode volume $V_{m}$ and large quality factor $Q$ \cite{braginsky89,vernooy98a}, an atom interacting with the evanescent field of a WGM can be well into
the regime of strong coupling with projected values for the critical photon $%
n_{0}$ and atom $N_{0}$ numbers $(n_{0},N_{0})\simeq (2\times
10^{-5},10^{-6})$ \cite{spillane05}, respectively, significantly beyond
current \cite{miller05} and projected \cite{hood01,spillane05} capabilities for
cavity QED with Fabry-Perot cavities (Fig. \ref{cQED}(c)).

The fabrication techniques pioneered in Refs. \cite{armani03,spillane03}
lend themselves to the integration of many microtoroidal resonators to form
optical networks, as illustrated in (a) and (b) of the figure. Part (a)
shows a photograph of silicon chip with a linear array of microtoroidal
resonators within a UHV apparatus \cite{dayan08}. The toroids appear as
small scattering centers on a silicon chip that runs vertically down the
center of the picture. The black arrows point to a horizontal SiO$_{2}$
fiber taper for coupling light to and from one resonator. Part (b) is a SEM
micrograph of an array of microtoroidal resonators showing toroids of fused
silica on silicon supports \cite{armani03}.

These resonators have the capability for input-output coupling with small
parasitic loss \cite{spillane03} for the configuration shown in (d), which
is a photograph of an individual toroid and fiber coupler from (a) \cite%
{dayan08}. Quality factors $Q=4\times 10^{8}$ have been realized at $\lambda
=1550$ nm\ and $Q\simeq 10^{8}$ at $\lambda =850$ nm, with good prospects
for improvement to $Q\sim 10^{10}$ \cite{spillane05}. For these parameters,
the efficiency $\epsilon $ for coupling quantum fields into and out of the
resonator could approach $\epsilon \sim 0.99-0.999$, while still remaining
firmly in the regime of strong coupling \cite{spillane05}. Such high
efficiency is critical for the realization of complex quantum networks,
including for the distribution and processing of quantum information \cite%
{briegel00,cirac97,duan04} and for investigations of the association between quantum many-body systems and quantum networks \cite{qmb,acin07}.

The initial step in this quest was the demonstration of strong coupling
between individual atoms and the field of a microtoroidal resonator in Ref.
\cite{aoki06}. More recently, nonclassical fields have been generated from
the interaction of single atoms with a microtoroidal resonator by way of a
`photon turnstile' for which a single atom dynamically regulates the
transport of photons one-by-one through the microtoroidal resonator \cite%
{dayan08}, as illustrated in (d). Only single photons can be transmitted in
the forward direction (to the left in the figure), with excess photons $n>1$
dynamically rerouted to the backward direction.

\begin{center}
\includegraphics[width=15cm]{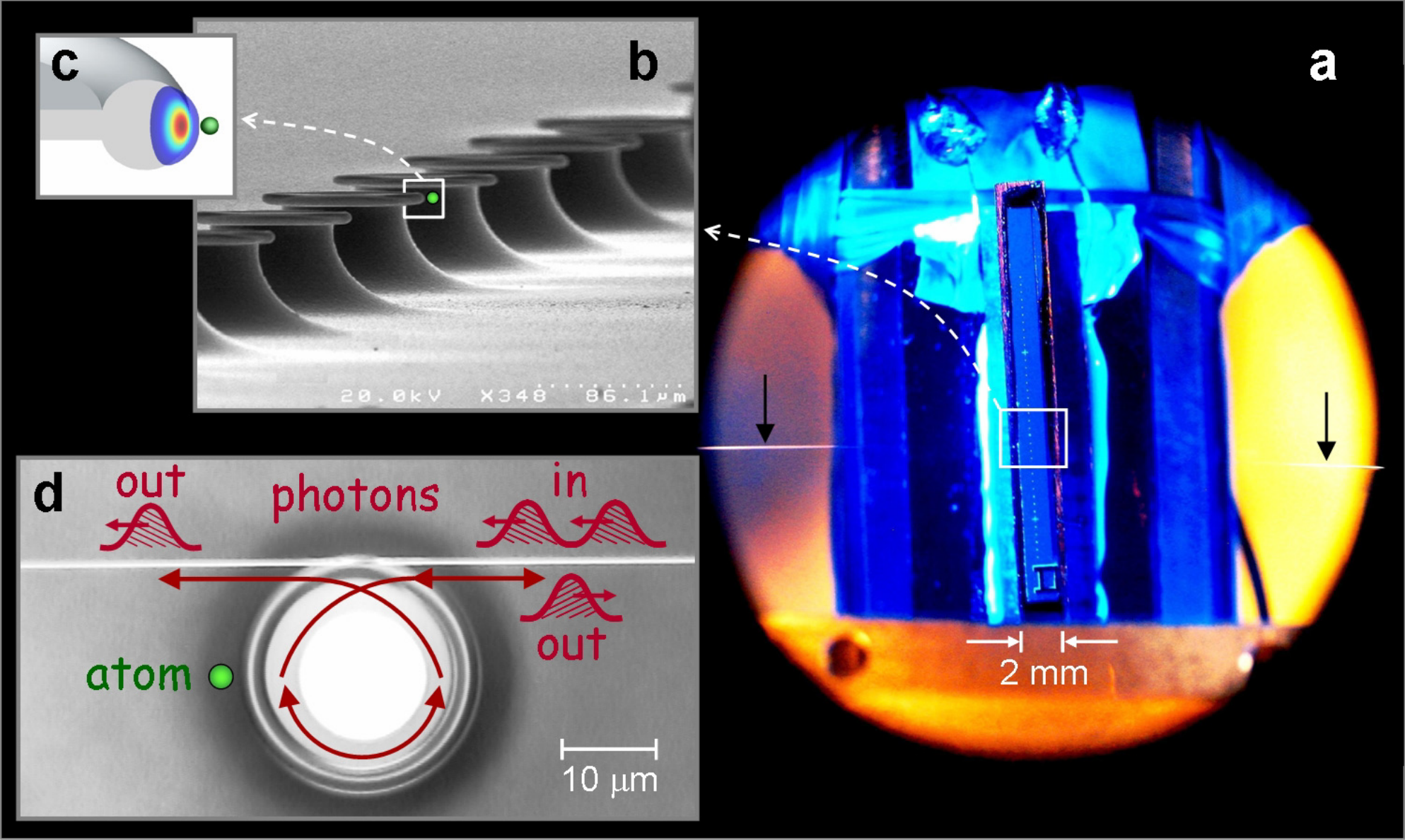}
\end{center}

\pagebreak

\section{Box 3 - Writing and reading single atomic excitations}

The DLCZ protocol \cite{duan01} is based upon ensembles of $N_{a}$ identical
atoms with a $\Lambda $-level configuration as shown in the figure. The
metastable lower states $\left\vert g\right\rangle $ and $\left\vert
s\right\rangle $ can be, e.g., atomic hyperfine states of the electronic
ground level to ensure a long lifetime for coherence. As illustrated in (a),
all atoms are initially prepared in state $\left\vert g\right\rangle $ with
no excitation, namely $\left\vert 0_{a}\right\rangle \equiv \otimes
_{i}^{N_{a}}\left\vert g\right\rangle _{i}$, and a weak off-resonant `write'
pulse is then sent through the ensemble. There results a small amplitude $%
\sqrt{p}$ for one of the $N_{a}$ atoms to be transferred from $\left\vert
g\right\rangle $ to $\left\vert s\right\rangle $ and to emit a photon into
the forward-scattered optical mode (designated as field $1$) with frequency
and/or polarization distinct from the write field.

Of course for small excitation probability $p\ll 1$, mostly nothing happens
as a result of the writing pulse, so that the resulting state $\left\vert
\phi _{a,1}\right\rangle $ for the atomic ensemble and field $1$ in the
ideal case is given by%
\begin{equation}
\left\vert \phi _{a,1}\right\rangle =\left\vert 0_{a}\right\rangle
\left\vert 0_{1}\right\rangle +e^{i\beta }\sqrt{p}\,\left\vert
1_{a}\right\rangle \left\vert 1_{1}\right\rangle +O(p)\text{,}  \tag{1}
\label{phia1}
\end{equation}%
where $\left\vert n_{1}\right\rangle $ is the state of the
forward-propagating field $1$ with $n_{1}$ photons and the phase $\beta $ is
determined by the write pulse and field $1$ propagation phases. The atomic
state $\left\vert 1_{a}\right\rangle $ in Eq. \ref{phia1} is a collective
(entangled) state with one excitation shared symmetrically among the $N_{a}$
atoms (i.e., one `spin flip'), where in the ideal case \cite{duan01}%
\begin{equation}
\left\vert 1_{a}\right\rangle =\frac{1}{\sqrt{N_{a}}}\sum_{i=1}^{N_{a}}\left%
\vert g\right\rangle _{1}\cdots \left\vert s\right\rangle _{i}\cdots
\left\vert g\right\rangle _{N_{a}}\text{.}  \tag{2}  \label{1a}
\end{equation}

Field $1$ is directed to a single-photon detector, where a detection event
is recorded with probability $\varpropto p$. Such an event for field $1$
heralds that a single excitation (or spin flip $\left\vert g\right\rangle
\rightarrow \left\vert s\right\rangle $) has been created and stored in the
atomic ensemble in the state $\left\vert 1_{a}\right\rangle $ with high
probability. Higher-order processes with multiple atomic and field $1$
excitations are also possible and ideally occur to lowest order with
probability $p^{2}$.

After a user-defined delay (subject to the finite lifetime of the quantum
memory), the collective atomic excitation $\left\vert 1_{a}\right\rangle $
can be efficiently converted to a propagating beam (designated as field $2$)
by way of a strong `read' pulse as in (b), where in the ideal case, there is
a one-to-one transformation of atomic to field excitation, $\left\vert
1_{a}\right\rangle \rightarrow \left\vert 1_{2}\right\rangle $. For the case
of resonance with the $\left\vert s\right\rangle \rightarrow \left\vert
e\right\rangle $ transition, the reading process utilizes the phenomenon of
electromagnetically induced transparency \cite{lukin-review,fleischhauer-review,harris97}.

\begin{center}
\includegraphics[width=12cm]{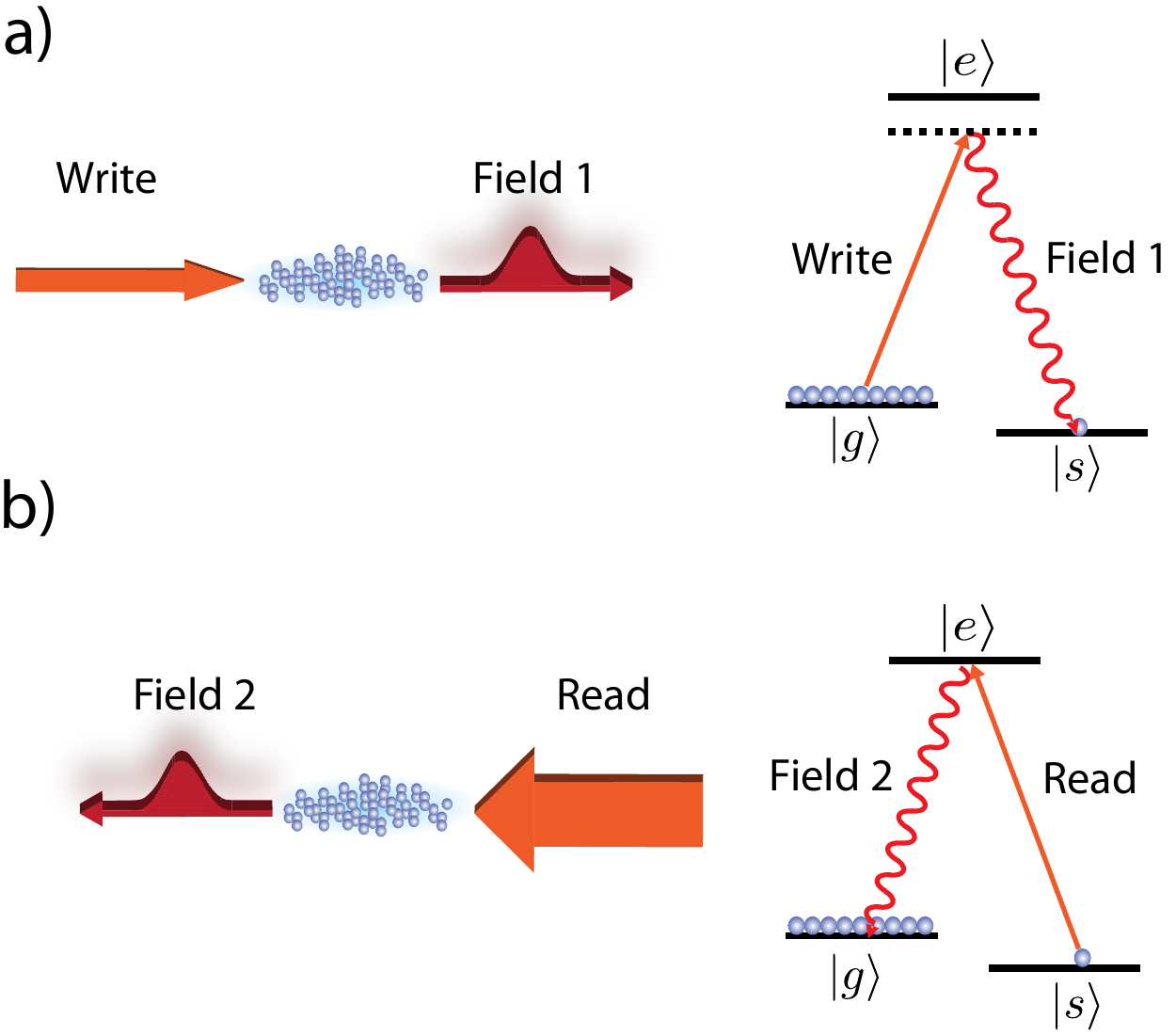}
\end{center}

\end{document}